\documentstyle[cjaa209,twoside,epsf]{article}
\input{psfig.sty}

\begin{document}
\title{Possible Streams of the Globular Clusters in the Galaxy}

\author{Shuang Gao $^{1}$ , Bi-Wei Jiang $^{1~\star}$ , Yong-Heng Zhao $^{2}$ }
\inst{
    $^1$ Department of Astronomy, Beijing Normal University, Beijing
    100875\\
    $^2$ National Astronomical Observatories, Chinese Academy of Sciences, Beijing 100012}
\email{bjiang@bnu.edu.cn}

\markboth{S. Gao, B.W. Jiang and Y.H.Zhao} { Possible Streams of the
Globular Clusters } \pagestyle{myheadings}
\date{Received  ; accepted }
\baselineskip=18pt

\begin{abstract}
This paper aims to retrieve the ghost streams under the
pre-assumption that all the globular clusters in our Galaxy were
formed in the very early merge events. The results are based on two
speculations: that the specific energy and angular momentum of the
globular clusters after merge are not changed in process of
evolution and that the globular clusters with common origin would
stay in the same orbit plane as parent galaxy. In addition, taking
into account the apo-galacticum distance of the orbits,  five
possible streams were suggested with a significant confidence. The
number of streams is consistent with previous results. Three of the
four well established members of the Sagittarius stream were found
to be in one of our streams. Several other globular clusters in our
result were also thought to come from accretion by previous
research. Furthermore, the orbital parameters of the streams are
derived, which provide a way to testify whether these streams are
true with the help of the accurate measurement of proper motions of
the globular clusters. 
\end{abstract}

\section{INTRODUCTION}
The globular cluster (GC hereafter), as the oldest star group in the
universe, has been a target that astrophysics has paid close
attention to all the time. The near-field (Galaxy) cosmology has
contacted closely with far-field cosmology by the Galaxy's GCs,
and the origin of GCs  has become an important component
of studying the Galaxy's formation and evolution (Freeman \&
Bland-Hawthorn 2002).

The earliest view on the Galactic formation is that the halo was
created during a rapid collapse of a relatively uniform, isolated
protogalactic cloud (Eggen, Lynden-bell \& Sandage 1962). Though
this view was challenged (Searle \& Zinn 1978), the theory that
GC originate from collapse of the Galactic molecular
cloud is stilled insisted by some groups. Harris \& Pudritz (1994)
argued that GCs follow a power-law number distribution
by mass, $N \sim M^{-1.7}$, which is virtually independent of
environment from the study of the connection between protogalactic
and present-day cluster formation by examining the GCLF (globular
cluster luminosity function) of several large galaxies in addition
to the Milky Way.

However, within the scenario of Cold Dark Matter (CDM) cosmology,
the universe is built up hierarchically, i.e. the small structure
forms early and the large structure forms lately. To our Milky Way
galaxy, such scenario fit the view of Searle \& Zinn (1978) that the
Galactic halo was built up over an extended period from independent
fragments. The discovery of the Sagittarius dwarf galaxy being
accreted by the Galaxy supports the view that the halo of the Galaxy
has been built up at least partially by the accretion of similar
dwarf galaxies (Ibata et al. 1994).

Schweizer (1987) first suspected that GCs were formed in mergers.
Later, Ashman \& Zepf (1992) predicted that the Hubble Space
Telescope would reveal young GCs which were later discovered in
disturbed or interacting galaxies such as NGC 1275 (Holtzman et al.
1992), NGC7252 (Whitmore et al. 1993) and the Antennae (Whitmore \&
Schweizer 1995). All these observations support the view that the
GCs could be formed in mergers. In our Galaxy, the proof continues.
From the Hertzsprung-Russell diagram, Lee et al. (1999) identified
multiple stellar populations traced by wide red giant branches in
the most massive GC $\omega$ Centauri with different
ages and argued that $\omega$ Centauri was once part of a more
massive system which merged with the Milky Way. Indeed, Lynden-Bell
\& Lynden-Bell (1995) traced the ghostly streams of the outer-halo
GCs and satellite galaxies starting from the opinion
that they were all formed in early merge events.

Recently, there are several efforts to estimate the number of merges
by comparing the characteristics of halo stars or GCs in the Galaxy
with those in extra-galaxies. In 1996, Unavane et al (1996) examined
the fraction of stars in the halo which have colors consistent with
a metal-poor, intermediate-age population matching those typically
observed in Local Group dwarf spheroidal galaxies. They concluded
that the star counts imply an upper limit of $\sim$60 merges with
low-luminosity dwarf spheroidal, or $\leq$6 mergers with more
luminous objects. In 2000, van den Bergh (2000), based on the survey
of young blue halo stars,  estimated that the total amount of
captured stellar material in the halo is equivalent to 3--7
Sagittarius dwarfs. In 2004, Mackey \& Gilmore (2004) compared in
details the properties of GCs between the old halo, young halo and
bulge$/$disk subsystems, and the GCs in LMC, SMC and Fornax and
Sagittarius dwarf spheroidal galaxies. They estimated the Galaxy may
have experienced approximately seven merger events with
cluster-bearing galaxies during its lifetime.

In this paper, we try to trace the streams, i.e. the satellite
galaxies, of the all the GCs in the halo from their kinetics other
than the properties.  The analysis presumes that some big satellite
galaxies had several GCs which kept their specific energy and
angular momentum and which moved in the same orbital plane after
merging with our Galaxy. We briefly describe our methods in Section
3 and discuss our results for the possible streams in Section 4.

\section{DATA}
All the parameters of GCs are taken from the web version of "A
Catalog of Parameters for GCs in the Milky Way"
(Harris 1996).  As Harris (2003) had critically selected the best
available measurements for each of the quantities included, the
quality of the measurements can be regarded at least one of the
best. In addition, the electronic version is available through web
and updated regularly, which guarantees the data being the latest.

The catalog is consisted of three parts, including positions,
photometric parameters and kinetic parameters respectively. It lists
the cluster identifications, positions, integrated magnitudes,
colors, morphology parameters, metallicities, radial velocities, and
structural parameters. We make use of the parameters such as radial
velocities, positions and distances. However, the necessary
parameters are not fully available for 3 GCs among the 150 GCs in
the Harris catalog because they are faint, inside the Galactic
plane, or lately found. So our analysis is directed to the 147 GCs
for which the needed parameters are complete without consideration
of the errors of the parameters.

\section{METHOD AND CALCULATION}

Two major criteria are set to sort the common origin of GCs. The
first criterion is that the GCs of common origin should have the
same specific energy and angular momentum. The second one is that
their orbits should lie within the same plane. At the beginning, all
the GCs are treated equally nevertheless they are different in some
characteristics such as metallicities or morphologies of horizonal
branches which are usually used in classification. Specifically,
they were all assumed to come from the accreted galaxies even if
some of them were originally produced in the Galaxy. However, the
final result from our analysis will screen out the members which
might be Galactic natives.

\subsection{Hypotheses}

To proceed our analysis, there are several strong hypotheses as
following which are essentially the same as those by Lynden-Bell \&
Lynden-Bell (1995).

\begin{itemize}
    \item Sufficiently large satellite galaxies have GCs
in their own halos. Such GCs would be pulled off if the satellite
galaxies were tidally torn. Indeed, it's this hypothesis that
motivated us to search for the Galactic GCs of common parent
galaxies, though the origin of the GCs in the large satellite
galaxies is another question.
    \item In tidal tearing, the objects
torn off pursue orbits in the same plane through the Galactic center
as their parent's orbit.
    \item The objects were torn off at
one or several close passages during a period in which the energy
and angular momentum of the progenitor's orbit did not change very
much.
    \item The initial orbits of the objects torn off had
    approximately the same specific energy and specific angular
    momentum as the progenitor.
    \item The gravitational potential in the Galactic halo has
    changed only slowly since that time and may now be taken to
    have a known form.
\end{itemize}

\subsection{Specific Energy and Angular Momentum}
As mentioned in the beginning of this section, the first criterion
of the GCs from the same parent galaxy is their equal specific
energy and angular momentum of the GCs. The specific energy $E$ and
the angular momentum $h$ are defined to be the total energy and
angular momentum per unit mass respectively. The equation of
specific energy is as follows:
                \begin{equation}\label{equ:er}
                   E_r=\frac{1}{2}v^2_r-\psi=E-\frac{1}{2}h^2
                   r^{-2} ,
                \end{equation}
where $E_r$ is the radial component of total energy, $r$, $\psi$ and
$v_r$ are the Galactocentric distance, gravitational potential and
Galactocentric radial velocity respectively.

According to equation (\ref{equ:er}), the GCs with the same specific
energy and angular momentum can be found out, if knowing the
gravitational potential of the halo, $\psi$, Galactocentric
distance, $r$, and the velocity $v_r$, because the GCs of the same
origin, which have the same specific energy and angular momentum,
are expected to lie on the same line delineated by equation
(\ref{equ:er}) with the gradient $-h^2/2$ and intercept $E$. Thus,
it is necessary to determine $\psi$, $r$ and $v_r$ to proceed.
 \begin{enumerate}
\item First, the gravitational potential $\psi$ is calculated.
Through the rotation curve of the Galaxy, we have the law that the
density in the Galactic halo vary with Galactocentric distance, thus
have the equation of the gravitational potential. Assuming the
density is positive everywhere and falls like $r^{-2}$ for $r\ll
r_h$, but like $r^{-5}$ for $r\gg r_h$, the form of the potential
can be given by
    \begin{equation}\label{}
        \psi=V_{0}^{2}\ln(\frac{\sqrt{1+S^2}+1}{S}) ,
    \end{equation}
where $V^2_0=220{\rm ~km~s^{-1}}$ is the circular velocity of the
Sun, $S=r/r_h$ and  $r_h=80{\rm kpc}$.  Thus, for a given $r$ which
is the Galactocentric distance, the potential can be calculated.

\item The Galactocentric distance $r$ can be calculated from the solar
distance and the Galactic coordinates from simple trigonal relation.

\item From the distant halo, the difference in direction of the
Sun and Galactic center can often be neglected, so the radial
velocity in the line of sight,${\it{v_l}}$ (after correction to the
Galactic center of rest), is approximately the radial velocity that
would be seen from the Galactic Center, $\it{v_r}$. Thus $\it{v_r}$
can be regarded as known. It should be kept in mind that this
approximation is virtually valid for the GCs distant from the
Galactic center and will bring about some uncertainty for those
relatively close to the Galactic center.
\end{enumerate}

\begin{figure}
   \begin{center}
   \mbox{\epsfxsize=\textwidth\epsfysize=\textwidth\epsfbox{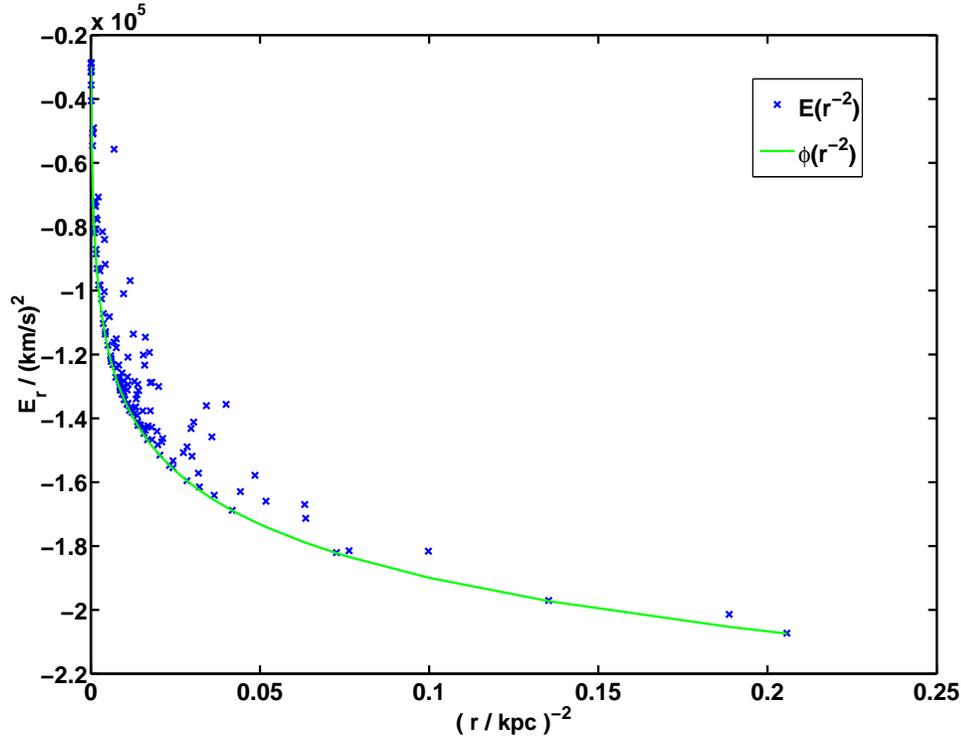}}
   \caption{$E_r=\frac{1}{2}v^2_r-\psi$ plotted against $r^{-2}$. Members of one stream
            should lie on one straight line in the diagram.}
   \label{Fig:E-r map}
   \end{center}
\end{figure}

From the available data, 147 of all the 150 GCs have been measured
in $v^2_r$ and $r$. The diagram of their radial energy $E_r$ vs
Galactocentric distance $r$ is shown in Fig.~\ref{Fig:E-r map},
where the solid curve describes the variation of the potential
$\psi$ with $r$.

To determine the straight lines on Fig.~\ref{Fig:E-r map} through
which some of the diffuse 147 points pass, a method called "Hough
Transform (hereafter HT)" is adopted. Because there are 147 points
on Fig.~\ref{Fig:E-r map}, the number of possible lines is
$C^{2}_{147}$ by connecting any two of the points. The HT method
essentially calculates all the slopes and intercepts of these
$C^{2}_{147}$ lines and finds out those lines which have close
slopes and intercepts, therefore the points making these lines
should have similar specific energy and angular momentum.

\begin{figure}
\hspace*{10mm}
    \mbox{\epsfxsize=0.5\textwidth\epsfysize=0.5\textwidth\epsfbox{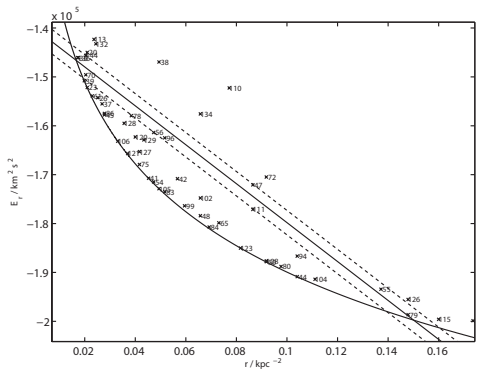}}
    \caption{One line in the radial energy diagram.
    There are 7 points between the upper line and lower line. These
    7 objects is classified as Stream no. 16 in Table~\ref{tab:21 streams}.
    The cross denotes a GC and the digit beside labels
    the sequence number of the corresponding GC in the Harris table. }
    \label{Fig:s16}
\end{figure}

The analysis showed that the gathering of the slopes of lines were
relatively concentrated while that of the intercepts were
scattering, thus grouping the specific energy and angular momentum
considered the error in the range of intercepts. As an example shown
in Fig.~\ref{Fig:s16},  the $1 \sigma$ errors are plotted with the
upper line and lower line being the error bounds. Consequently, the
points within $1 \sigma$ error were regarded to have the same
specific energy and angular momentum, i.e. passing  the first
criterion examination. This analysis of the 147 points yielded 21
lines, which means there are 21 groups with the same $E$ and $h^2$.
The names and number of the members in each group are shown in
Table\ref{tab:21 streams}. The streams in Table~\ref{tab:21 streams}
are sorted by the intercepts of the lines decreasing from top to
bottom, which is indeed the specific total energy $E$.


It can be seen from Table \ref{tab:21 streams} that the number of
members of each group ranges from 3 to 14. Indeed, more than 3
points on a line is required to make a group.  As an example, one of
the 21 lines is displayed in Fig.~\ref{Fig:s16}, where the points in
the same group are between the upper line and lower line.

\subsection{Polar-path Maps}

The objects satisfying the second criterion, i.e. moving in the same
orbits, is selected through the Lambert equal-area projection.
Different from the traditional projection, in Lambert projection,
the x-y plane's transform equations are following:
    \begin{eqnarray}
      x' &=& \frac{x}{\sqrt{1+z}} , \\
      y' &=& \frac{y}{\sqrt{1+z}} ,
    \end{eqnarray}
where $(x',y')$ are coordinates transformed from the x-y plane.
After such transformation, the size of projected area is
proportional to and thus retains the original relative sizes, which
is very important in present work because only its position is known
for a given GC.

The poles of the countless orbits resulted from the limited
information (i.e. the position of the GC and that the Galactic
center is within the orbital plane) consist a set which is called
polar-path. The equation of the polar-path is described in the
Cartesian coordinates by:
    \begin{equation}\label{equ:polar-path-right}
    x^{2}(1+\frac{\cos^2 l}{\tan^2 b})+y^{2}(1+\frac{\sin^2 l}{\tan^2
b})+2xy\frac{\sin l \cos l}{\tan^2 b}=1 ,
    \end{equation}
    where $(l,b)$ are the Galactic coordinates of the given GC.


The search for the common orbits is converted to search for the
intersection points of the polar paths of the GCs. For every of the
21 groups in Table~\ref{tab:21 streams}, the polar-path is plotted
for each GC in the projected plane. Whether the polar-paths have
intersections, which was judged by eyes, demonstrates whether they
have the same orbit. It is found that all or part of the members
have intersections in eight groups, i.e. they have the same orbits.
while no more than 3 members have intersections in thirteen groups
which are regarded not to move in the same orbit.

\begin{figure}
    \begin{center}
    \mbox{\epsfxsize=0.8\textwidth\epsfysize=0.8\textwidth\epsfbox{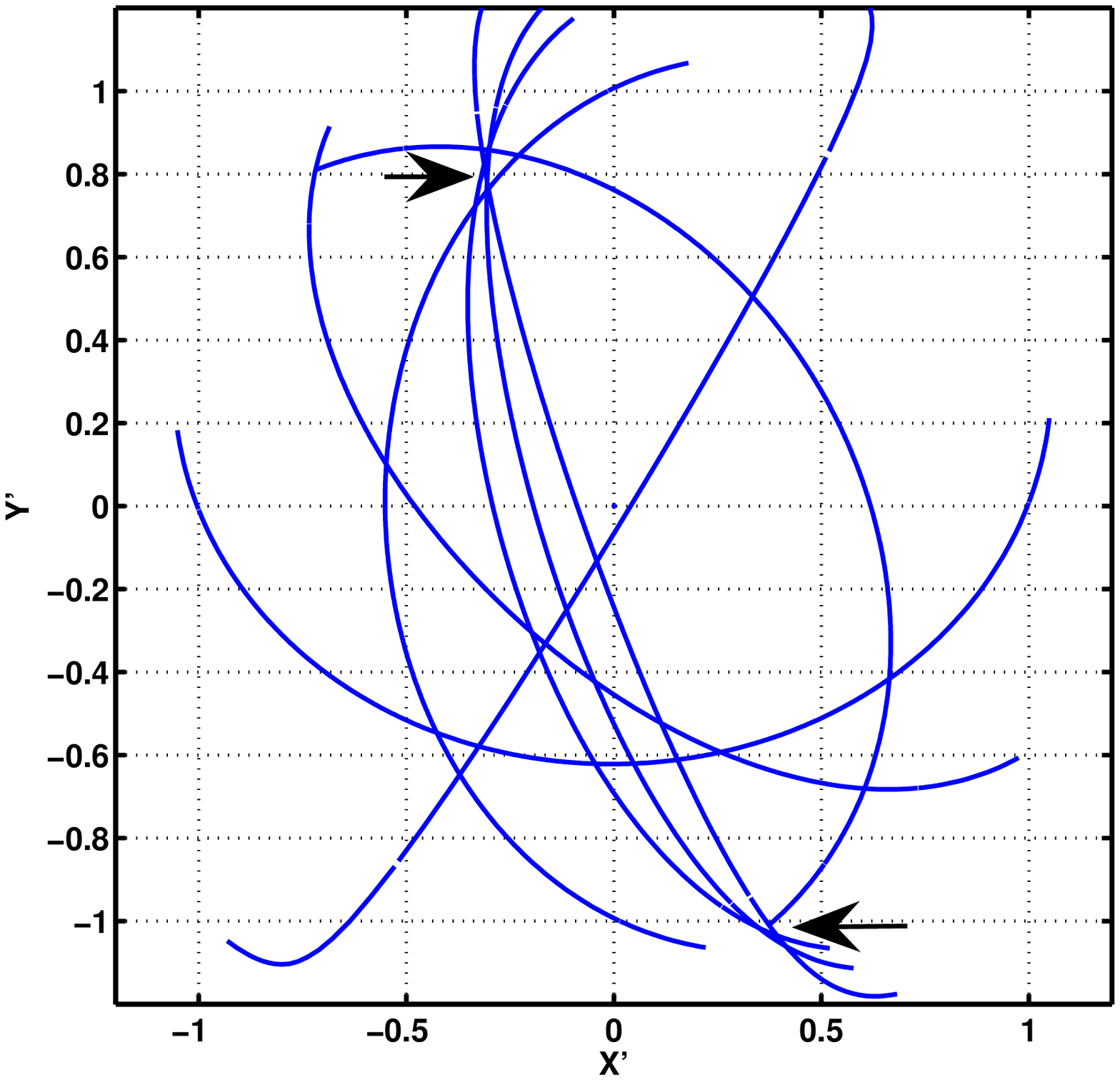}}
    \mbox{\epsfxsize=0.8\textwidth\epsfysize=0.8\textwidth\epsfbox{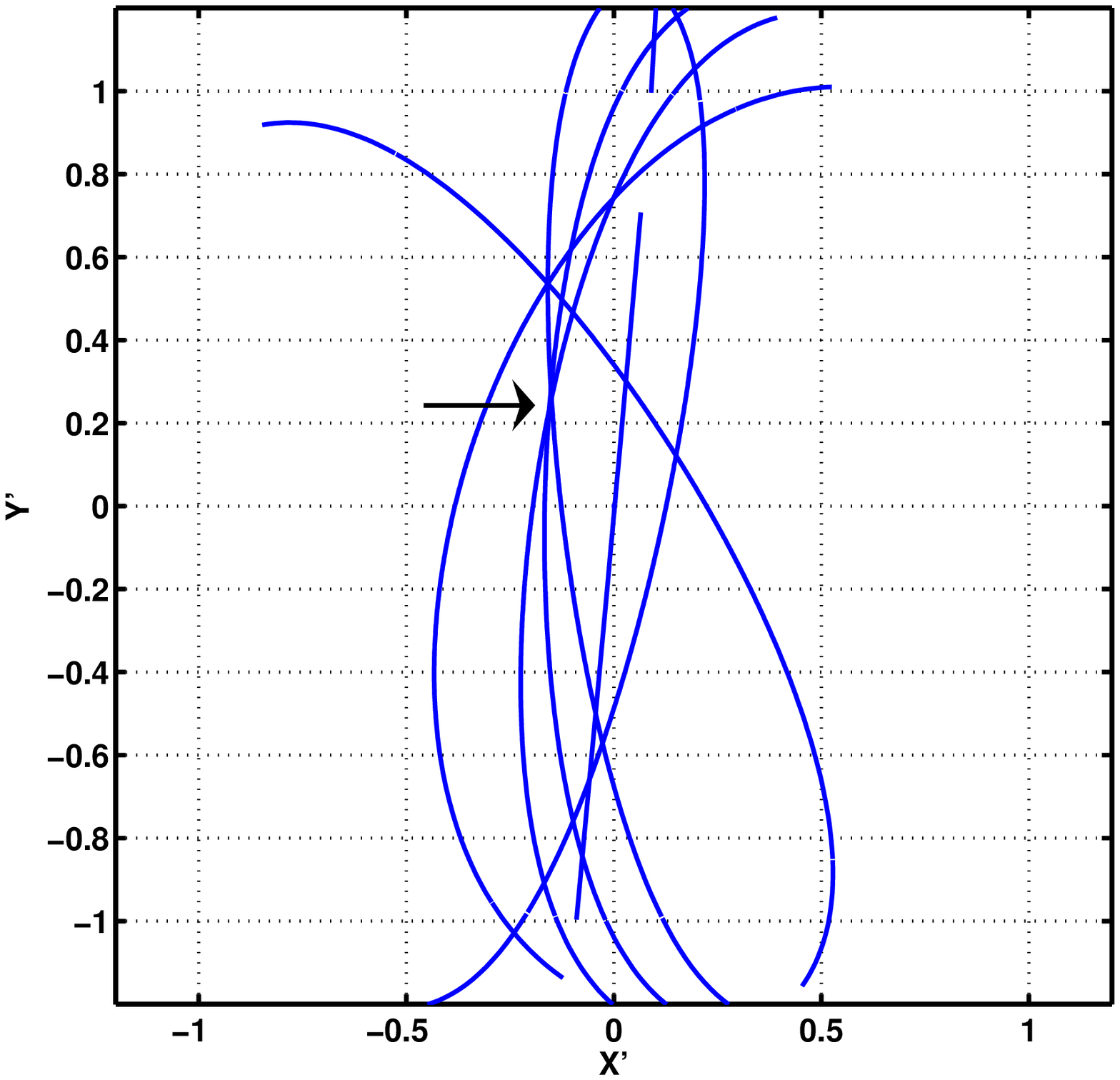}}
    \caption{Polar paths of two groups of globular clusters.
       The left: the members of stream no. 8 streams in
                Table\ref{tab:21 streams}. The right: the members of
                stream no. 16 in
                Table\ref{tab:21 streams}. Each path is a great circle perpendicular
                to the galactocentric vector to the globular cluster,
                $X'^2+Y'^2=1$ (see text for details)
                is the Galactic plane. $X'^2+Y'^2<1$ is the northern Galactic sky.
                Multiple intersections of polar paths means possible stream poles. The
                arrow in the diagram points out these poles. }
    \label{Fig:polar}
    \end{center}
\end{figure}

These eight groups for which some members are found to have
polar-path intersections are listed in Table~\ref{tab:8 streams}, as
well as the NGC serial numbers of the members with common poles of
orbital planes (some members in Table~\ref{tab:21 streams} are
dropped due to lack of intersection with others). As examples, the
polar paths of the members in two groups are shown in Figure
\ref{Fig:polar}, where the left panel indicates a successful one,
the eighth group in the 21 streams in Table~\ref{tab:21 streams},
and the right panel displays the failed search, number 16 in
Table~\ref{tab:21 streams}.

\subsection{Error Analysis}

During the process to search for the groups, some errors are induced
from approximations or methods. Firstly, the Galactocentric radial
velocity is approximated by the line-of-sight velocity. This
approximation is true only for the objects which are very distant
from the Galactic center where the position difference between the
Sun and the Galactic center has little effect on the velocity
difference. It brings about significant error in the case of nearby
GCs. Secondly, the error is induced from applying the HT method to
pick out the points on one line. These errors have been utilized to
restrict the ranges of selecting points, and have been plotted on
Fig.\ref{Fig:s16} (dashed lines). Because there are uncertainties in
the parameters of velocity and distance of GCs, we suffered some
dispersion in the transformed slope-intercept diagram (see Section
3.2 for details) of the points. This means that the members in the
same group may have some scattering in the specific energy and
angular momentum. Thirdly, the common poles from the Lambert's
equal-area projection are not exactly at one point, but with some
range. Whether the GCs have the same orbital pole is judged by eyes,
with the difference in $X'$ and $Y'$ less than 0.1 or so.

\section{RESULT AND DISCUSSION}

\subsection{The Orbital Parameters}

From the analysis of the specific energy and angular momentum, and
the polar path of the GCs, we found 8 groups, with the members in
each group possibly having common satellite origin.  Furthermore,
based on the position parameters of the members and the pole
position as the intersections in Fig.~\ref{Fig:polar} in each of the
eight streams, the equation of the normal to each orbital plane was
solved. The orbital parameters of the 8 groups are shown in Table
\ref{tab:results}, where the stream number (first appearing in
Table~\ref{tab:21 streams}), the apo-galacticum, peri-galacticum,
eccentricity, specific angular momentum, specific total energy,
period and coefficients of the orbital equation are listed from left
to right. The ambiguity of the direction of the normal to the
orbital plane brings about two combinations of the coefficients.
That's why the last two columns have two combinations. This
degeneracy can be relieved by estimating the direction of the
angular momentum, which will be carried out in future more detailed
work.


It can be seen from Table~\ref{tab:results} that the streams no. 18,
19, and 21 have their apo-galacticum less than 8.5kpc, the adopted solar
distance to the Galactic center. For the members in these
 streams, the approximation of the solar radial velocity to
 the Galactic radial velocity could bring about large error
 in calculating the specific energy and angular momentum.
 Moreover, the GCs with the orbit so close to the
Galactic center should have been disrupted due to the strong tidal
force if they had been accreted long before. Another disfavor of these three
objects is that one object (NGC 5904 and NGC 6558) appears in two streams,
which indicates mediocre quality in grouping. Therefore, these three
streams should not be the ghost streams, which are separated by a line in
Table~\ref{tab:results} from the candidate streams. Finally, the five
streams, specifically no. 5, 6, 8, 9 and 10, are possible streams of
the GCs in the Galaxy from the kinetic point of view.

The eccentricity of the GCs's orbits has been discussed widely
because it is related to their origins as well as the evolution.
Statler(1988) suggested that the orbital eccentricity of GCs is
decreasing in the process of dissipative through numerical
simulation method. Gnedin(2006) derived the average eccentricity of
surviving model clusters to concentrate between 0.4 and 0.8. This is
consistent with the early result by Ninkovi\'{c} (1983).  In
Table.\ref{tab:results}, the eccentricities of possible streams are
between $0.5$ to $0.8$, which agrees well with the high probability
eccentricities of GCs. The result indirectly confirms our
calculation.

\subsection{Number of Streams}

From our analysis, there are five streams for the origin of about
20\% GCs in the Galaxy, which means there had been five relatively
large and cluster-bearing satellite galaxies being accreted in early
times. This number is well consistent with others. Unavane et al.
(1996) concluded $\leq$ 6 mergers with luminous galaxies. van den
Bergh (2000) estimated the total amount of captured stellar material
equivalent to 3-7 Sagittarius dwarfs. Mackey \& Gilmore (2004)
estimated the Galaxy may have experienced approximately seven merger
events with cluster-bearing galaxies during its lifetime. All these
results agree reasonably well with ours.

\subsection{The Sagittarius Stream}

As the best established stream, the Sagittarius stream has been
investigated in most details.  Consequently, some GCs have been
suggested as the members of the Sagittarius stream. The discovers
Ibata et al. (1997) suggested NGC 6715 (also known as M54), Ter 8,
Ter 7 and Arp 2 being associated with the Sagittarius stream.
Bellazzini et al (2003) added Pal 12, NGC 4147. The 2MASS
color-magnitude diagram (Bellazzini et al. 2004) enhanced their
belief that the two GCs are the members. Pal. 12
(Mart\'{\i}nez-Delgado et al. 2002) was also proposed ro be a
member. Although numerous additional members have been postulated,
there are only four well-established members as suggested by Ibata
et al. (1997). From Table~\ref{tab:8 streams}, stream no.6 includes
three of the four established members, i.e. NGC 6715, Ter 8 and Ter
7, which is in good coincidence with previous results. The left
member Arp 2 does not appear in this group, which may be caused by
the error in our analysis or that it is not the member from the
kinetic point of view. The other postulated members neither appear
in stream no.6. In addition, the other member in stream no.6 is
IC-1276 which is in the Galactic bulge (Barbuy et al. 1998), and
thus possibly is not the member. We conclude that stream no.6 is at
least part of the Sagittarius stream.

\subsection{Other Members}

In addition to the members of the Sagittarius stream, several more
GCs in our streams also appear in previous identifications.  NGC
7492, NGC 7089, NGC 6809 and NGC 5904 which appear in stream no. 5,
10, 18 and 19 (though no. 18 and 19 may not be the stream due to
their close orbit to the Galactic center) were suggested to be
similar to GCs in extragalaxies by Mackey \& Gilmore (2004). Pal 5
which appears in stream no. 8 was considered to lie in a stream from
the SDSS-II data (Odenkirchen et al. 2003). Our result confirmed
some GCs speculated to originate externally. However, the exact
streams are difficult to configure. Besides, some of the GCs in our
results have not been found to be of external origin by other
methods. Because the method in our analysis suffers some uncertainty
(especially lack of the accurate measurement of proper motions) and
that the judgement only from the properties of GCs is not decisive,
further investigations are needed to clarify the origins of the GCs
in the Galaxy.

\section{Summary} The present work agree with that the Galaxy had experienced
several merges. Our result classified the GCs of possible external
origin into 5 groups, which is well consistent with previous
results.  In particular, three of the four well established
Sagittarius members were classified into one stream. As previous
work identified the external origin of GCs mainly from their
observed properties such colors, metallicity or age,  our work
identified their origin from their kinematic properties. Since our
view is different from others, the consistent conclusion is
important.

Whether the stream found is true or imaginary needs further
investigations, even if the uncertainties mentioned above could all
be eliminated. This is because the features --- the same specific
angular momentum and the same orbit plane --- can not guarantee the
GCs originated from external galaxies, i.e., the GCs of the same
origin inside the Galaxy would also preserve such characteristics.
However, other properties will help to judge if our sort of groups
is right or wrong. For example, the accurate determination of the
GCs' proper motions will make it possible to calculate the three
dimensional velocities and thus the right orbits. Presently
available data and observational techniques can yet provide such
data. The launch of GAIA (Mignard 2005) will provide high precision
measurements of proper motions to Galactic objects. We expect then
the test of our results. \\

{\noindent \bf ACKNOWLEDGMENTS} We thank the anonymous referee who
provided a helpful critique of the manuscript. This research was
supported by NSFC under grant 10573022, and the Undergraduate
Research Foundation of Beijing Normal University.

    \begin{table}
    \caption{The intermediate results from the specific angular momentum and energy of all
    the 147 GCs, including 21 streams. The columns are: stream serial number, number of members
    in the stream, the members in the stream with the NGC serial number, or the name if not
    included in the NGC catalogue. }
    \label{tab:21 streams}
    \begin{center}\begin{tabular}{ccl}
      \hline
\\
    stream$\sharp$ & Counts& Members' ID(NGC) \\
\\
      \hline
\\
      1            & 5     &  6539  6453  6352  5824  7492  \\
      2            & 3     &  6712  5824  7492 \\
      3            & 7     &  6517  Lynga 7     6496  6229  5824  7492  Terzan 3  \\
      4            & 9     &  6362  6656  6749  5986  3201  6229  5824  7492  Terzan 3 \\
      5            &10     &  6544  Pal 10      6584  4590  6229  5824  7492  Terzan 3    6934  6981 \\
      6            & 7     &  IC 1276     4147  6715  Terzan 8    Terzan 7    6934  6981 \\
      7            & 6     &  Pal 8       6752  6101  1851  5466  5634 \\
      8            & 8     &  6517  Lynga 7     6254  1261  1904  Pal 5       IC 1257     Pal 12 \\
      9            & 5     &  6402  6535  6496  6205  6864  \\
     10            &14     &  6426  6864  5272  2808  7089  6838  6584  Pal 10      6362  6760  6656  6218  6540  6388  \\
     11            & 4     &  288   Terzan 12   6453  6539  \\
     12            & 8     &  288   7078  104   6838  6496  6535  6316  6712 \\
     13            & 9     &  6779  104   6838  5286  6584  Pal 10      6144  6341  362 \\
     14            & 8     &  5286  4372  Pal 10      6254  6496  6341  6171  6553 \\
     15            & 6     &  6441  6352  4372  7099  5897  Pal 11   \\
     16            & 7     &  Liller 1    6235  6144  6496  6356  Pal 11      5897  \\
     17            & 6     &  6356  5946  4833  6284  5139  6809 \\
     18            & 8     &  6440  6681  6809  5946  4833  6284  5139  5904 \\
     19            & 8     &  Terzan 1    6558  5946  4833  6284  5139  5904  6362 \\
     20            & 7     &  6254  6656  6496  6760  6544  6749  6333 \\
     21            & 6     &  6712  6453  6558  6355  6553  6171  \\
\\
    \hline
    \end{tabular}\end{center}
    \end{table}

    \begin{table}
    \caption{The final results include 8 groups in which part or all members
    have a common orbital poles. The columns are the same as
    Table~\ref{tab:21 streams}. The GCs which were regarded coming from accretion by
    other research are labeled and the corresponding references are shown at the footnotes. See the text for details.}

    \label{tab:8 streams}
    \begin{center}\begin{tabular}{ccl}
      \hline
\\
    Stream$\sharp$ & Counts& Members' ID(NGC) \\
\\
      \hline
\\
      5            & 3     & Pal-10~~~~~7492\footnotemark[3]~~~6934 \\
      6            & 4     & IC-1276~~~6715\footnotemark[1]~~~Ter-7\footnotemark[1]~~Ter-8\footnotemark[1]\\
      8            & 5     & 6517~~~~~~~6254~~~~Pal-5\footnotemark[2]~~~Pal-12~~~IC-1257 \\
      9            & 3     & 6402~~~~~~~6535~~~~6864 \\
     10            & 4     & 5272~~~~~~~7089\footnotemark[3]~~~6838~~~~~Pal-10 \\
     18            & 4     & 6440~~~~~~~6681~~~~6809\footnotemark[3]~~~~5904\footnotemark[3] \\
     19            & 4     & Ter-1~~~~~~6558~~~~5904\footnotemark[3]~~~~6284 \\
     21            & 5     & 6453~~~~~~~6558~~~~6355~~~~~6553~~~~~~6171 \\
\\
    \hline

    \end{tabular}\end{center}

\footnotemark[1]Ibata et al. 1997 \\
\footnotemark[2]Odenkirchen M. et al. 2003\\
\footnotemark[3]Mackey \& Gilmore  2004

    \end{table}

\begin{table}
    \caption{The results with the 8 orbital parameters and the quality of result.
             The $r_{max}$ and $r_{min}$ are the orbital apo-galacticum and peri-galacticum
             distances respectively, $e$ is the orbital eccentricity, $P$ is orbital period,  $a$, $b$ and $c$ are the parameters
             of the normal to orbital plane delineated by $ax+by+cz=0$.}
    \label{tab:results}
\begin{center}
\begin{tabular}{crccccrrc}
  \hline \\
  Stream$\sharp$     &     $\frac{r_{max}}{kpc}$     &     $\frac{r_{min}}{kpc}$
  &     $e$          &
        $\frac{10^{-3}h}{kpc\cdot km/s}$             & $\frac{-10^{-5}E}{(km/s)^2}$ &
        $\frac{P}{10^{8}yr}$                         &     $\frac{b}{a}$     &     $\frac{c}{a}$         \\ \\
  \hline
  5 & 29.88 & 5.50 & 0.69 & 2.2414 & 0.8 & 3.11 &  $\mp0.78$ & $0$         \\
  6 & 24.84 & 2.72 & 0.80 & 1.2604 & 0.9 & 2.76 &  $\mp0.39$ & $\mp2.31 $    \\
  8 & 19.15 & 4.00 & 0.65 & 1.5868 & 1.0 & 1.96 &  $\mp3.08$ & $\mp0.84 $   \\
  9 & 15.21 & 3.73 & 0.64 & 1.4142 & 1.1 & 1.45 &  $\mp2.34$ & $\pm0.29 $    \\
  10& 14.75 & 4.46 & 0.54 & 1.5868 & 1.1 & 1.51 &  $\mp0.71$ & $\mp0.02 $   \\
  \hline
  18&  6.84 & 1.21 & 0.70 & 0.5018 & 1.5 & 0.71 &  $\mp9.69$ & $\pm0.64 $    \\
  19&  6.73 & 1.42 & 0.65 & 0.5630 & 1.5 & 0.69 &  $\pm58.29$& $\pm1.03 $     \\
  21&  3.49 & 0.97 & 0.57 & 0.3552 & 1.8 & 0.35 &  $\mp64.38$& $\pm3.94 $    \\
  \hline
\end{tabular}\end{center}
\end{table}

\begin{thebibliography}
\baselineskip=12pt
\parskip=-3pt
\bibitem[1992]{ashm92}  Ashman K.M., Zepf S.E, 1992, \apj, 384, 50

\bibitem[1998]{barbuy98}Barbuy B., Ortolani S., Bica E., 1998,
\aas, 132, 333B

\bibitem[2004]{bell04}  Bellazzini M., Ibata R., Ferraro F., 2004,
In: F. Prada, D. Martinez, T.J. Mahoney, eds, ASP Conf. Ser. Vol.
327, Satellites and Tidal Streams, San Francisco: ASP, p. 220

\bibitem[2003]{bell03}  Bellazzini M., Ibata R., Ferraro F., Testa V.,
2003, A\&A, 405, 577

\bibitem[2000]{els62}   Eggen O.J., Lynden-Bell D., Sandage A.R., 1962, \apj, 136, 748 (ELS)

\bibitem[2002]{free02}  Freeman K., Bland-Hawthorn J., 2002, \araa, 40, 487

\bibitem[2006]{gnedin06}Gnedun Y., Prieto L., 2006, In: T. Richtler, et
al eds, Invited Review for Conference "Globular Clusters, Guide to
Galaxies", Chile: University of Concepcion

\bibitem[1994]{harr94}  Harris E, Pudritz E., 1994, \apj, 429, 177

\bibitem[1996]{harr96}  Harris E., 1996, \aj, 112, 1487

\bibitem[Harris(2003)]{Harris03}Harris, W. 2003, available at
{http://www.physics.mcmaster.ca/resources/globular.html} and {http://www.mporzio.astro.it/\~{}marco/gc/}

\bibitem[Holtzman et al.(1992)]{Holtzman92}Holtzman, J., Faber, S., Shaya, E. et al., 1992, \aj, 103, 691

\bibitem[1994]{ibata94} Ibata R., Gilmore G., Irwin M., 1994, \nat, 370, 194

\bibitem[1997]{ibata97} Ibata R., Wyse R., Gilmore, G. et al., 1997, \aj, 113,, 634

\bibitem[Lee et al.(1999)]{Lee99} Lee Y.W., Joo J.M., Sohn Y.J. et al., 1999, \nat, 402, 55

\bibitem[1995]{lynd95b} Lynden-Bell D., Lynden-Bell R.M., 1995, \mnras, 275, 425

\bibitem[2004]{mack04} Mackey A., Gilmore G., 2004, \mnras, 355, 504

\bibitem[2002]{mart02} Mart\'{\i}nez-Delgado D., Zinn R., Carrera R., et al., 2002, \apj, 573, L19

\bibitem[2005]{Mig05} Mignard F., 2005, In: Turon C., O'Flaherty K., Perryman M., eds,
the Observatoire de Paris-Meudon, The proceedings of the Gaia
Symposium "The Three-Dimensional Universe with Gaia", SP 576, 5,
Paris: ESA

\bibitem[1983]{Nin83} Ninkovi\'{c} S., 1983, Anstron. Nachr., 304,
305

\bibitem[Odenkirchen(2003)]{Odenkirchen03}Odenkirchen M., Grebel Eva K., Dehnen W., et al. 2003, \aj, 126, 2385O

\bibitem[Schweizer(1987)]{Schweizer87}Schweizer, F. 1987, in Nearly Normal Galaxies:
   From the Planck Time to the Present, ed. Faber, S., 18

\bibitem[1978]{sear78}  Searle L., Zinn R., 1978, \apj, 225, 357

\bibitem[1988]{stat88}  Statler S., 1988, \apj, 331, 71

\bibitem[1996]{unav96} Unavane M., Wyse R., Gilmore G., 1996, \mnras, 278, 727

\bibitem[1993]{berg00}  van den Bergh S. 2000, \apj, 530, 777

\bibitem[Whitmore et al.(1993)]{Whitmore93}Whitmore B., Schweizer F., Leitherer C.,
   et al., 1993, \aj, 106, 1354

\bibitem[Whitmore \& Schweizer(1995)]{Whitmore95}Whitmore B., Schweizer F. 1995, \aj, 109, 960

\end{thebibliography}
\end{document}